\documentclass[%
aps,
pre,
 amsmath,amssymb,
 superscriptaddress,
preprint,%
longbibliography
]{revtex4-1}

\usepackage{hyperref} 

\usepackage{graphicx}
\usepackage{dcolumn}
\usepackage{bm}
\usepackage{ulem}
\usepackage[utf8]{inputenc}
\usepackage[T1]{fontenc}
\usepackage{mathptmx}
\usepackage{textcomp} 
\usepackage{nicefrac}
\usepackage{xcolor}
\usepackage{gensymb}
\usepackage{multibib}
\usepackage{multirow}

\newcommand{\ve}[1]{\boldsymbol{#1}}
\usepackage{lineno}

\def\bibsection{\section{\refname}} 

\DeclareMathAlphabet{\pazocal}{OMS}{zplm}{m}{n}

\newcommand{\GW}{\ensuremath{G_{0}W_{0}}}

\newcommand{\exciting}{{\usefont{T1}{lmtt}{b}{n}exciting}~}
\newcommand{\ie}{{\it i.e.}, }

\begin{document}

\title{Hybrid Frenkel-Wannier excitons facilitate ultrafast energy transfer at a 2D-organic interface}

\author{Wiebke Bennecke} %
\address{I. Physikalisches Institut, Georg-August-Universit\"at G\"ottingen, Friedrich-Hund-Platz 1, 37077 G\"ottingen, Germany}

\author{Ignacio Gonzalez Oliva} %
\address{Physics Department and CSMB, Humboldt-Universität zu Berlin, 12489 Berlin, Germany}

\author{Jan Philipp Bange} %
\address{I. Physikalisches Institut, Georg-August-Universit\"at G\"ottingen, Friedrich-Hund-Platz 1, 37077 G\"ottingen, Germany}

\author{Paul Werner} %
\address{I. Physikalisches Institut, Georg-August-Universit\"at G\"ottingen, Friedrich-Hund-Platz 1, 37077 G\"ottingen, Germany}

\author{David Schmitt} %
\address{I. Physikalisches Institut, Georg-August-Universit\"at G\"ottingen, Friedrich-Hund-Platz 1, 37077 G\"ottingen, Germany}

\author{Marco Merboldt} %
\address{I. Physikalisches Institut, Georg-August-Universit\"at G\"ottingen, Friedrich-Hund-Platz 1, 37077 G\"ottingen, Germany}

\author{Anna M. Seiler} %
\address{I. Physikalisches Institut, Georg-August-Universit\"at G\"ottingen, Friedrich-Hund-Platz 1, 37077 G\"ottingen, Germany}

\author{Kenji Watanabe} %
\address{Research Center for Electronic and Optical Materials, National Institute for Materials Science, 1-1 Namiki, Tsukuba 305-0044, Japan}

\author{Takashi Taniguchi} %
\address{Research Center for Materials Nanoarchitectonics, National Institute for Materials Science,  1-1 Namiki, Tsukuba 305-0044, Japan}

\author{Daniel Steil} %
\address{I. Physikalisches Institut, Georg-August-Universit\"at G\"ottingen, Friedrich-Hund-Platz 1, 37077 G\"ottingen, Germany}

\author{R. Thomas Weitz} %
\address{I. Physikalisches Institut, Georg-August-Universit\"at G\"ottingen, Friedrich-Hund-Platz 1, 37077 G\"ottingen, Germany}
\address{International Center for Advanced Studies of Energy Conversion (ICASEC), University of Göttingen, Göttingen, Germany}

\author{Peter Puschnig}%
\address{Institute of Physics, NAWI Graz, University of Graz, 8010 Graz, Austria}%

\author{Claudia Draxl} %
\address{Physics Department and CSMB, Humboldt-Universität zu Berlin, 12489 Berlin, Germany}
\address{European Theoretical Spectroscopic Facility (ETSF)}

\author{G.~S.~Matthijs~Jansen} \email{gsmjansen@uni-goettingen.de}%
\address{I. Physikalisches Institut, Georg-August-Universit\"at G\"ottingen, Friedrich-Hund-Platz 1, 37077 G\"ottingen, Germany}

\author{Marcel Reutzel} \email{marcel.reutzel@phys.uni-goettingen.de}%
\address{I. Physikalisches Institut, Georg-August-Universit\"at G\"ottingen, Friedrich-Hund-Platz 1, 37077 G\"ottingen, Germany}

\author{Stefan Mathias} \email{smathias@uni-goettingen.de}%
\address{I. Physikalisches Institut, Georg-August-Universit\"at G\"ottingen, Friedrich-Hund-Platz 1, 37077 G\"ottingen, Germany}
\address{International Center for Advanced Studies of Energy Conversion (ICASEC), University of Göttingen, Göttingen, Germany}

\begin{abstract}

Two-dimensional transition metal dichalcogenides (TMDs) and organic semiconductors (OSCs) have emerged as promising material platforms for next-generation optoelectronic devices. The combination of both is predicted to yield emergent properties while retaining the advantages of their individual components. In OSCs the optoelectronic response is typically dominated by localized Frenkel-type excitons, whereas TMDs host delocalized Wannier-type excitons. However, much less is known about the spatial and electronic characteristics of excitons at hybrid TMD/OSC interfaces, which ultimately determine the possible energy and charge transfer mechanisms across the 2D-organic interface. Here, we use ultrafast momentum microscopy and many-body perturbation theory to elucidate a hybrid exciton at an TMD/OSC interface that forms via the ultrafast resonant Förster energy transfer process. We show that this hybrid exciton has both Frenkel- and Wannier-type contributions: Concomitant intra- and interlayer electron-hole transitions within the OSC layer and across the TMD/OSC interface, respectively, give rise to an exciton wavefunction with mixed Frenkel-Wannier character. By combining theory and experiment, our work provides previously inaccessible insights into the nature of hybrid excitons at TMD/OSC interfaces. It thus paves the way to a fundamental understanding of charge and energy transfer processes across 2D-organic heterostructures.

\end{abstract}

\maketitle

\noindent\textbf{Introduction}
\vspace{0.2cm}

Hybrid Frenkel-Wannier excitons are Coulomb-bound electron-hole pairs that join the unique optical properties of Frenkel excitons with the wavefunction delocalization of Wannier excitons~\cite{Agranovich11cr, Krumland21electronicstructure}. Such excitons have been predicted to occur at the interface of organic-inorganic heterostructures~\cite{Gonzalez22prm}, where they would mediate charge and energy transport~\cite{Homan17nanolett, Zhu18sciadv, Markeev22acsnano, Thompson23npj2d, Rijal20jpcl}. Hybrid Frenkel-Wannier excitons have also been proposed as a promising platform for studying many-body exciton physics~\cite{ulman_organic-2d_2021} as well as for nonlinear optical applications~\cite{Agranovich11cr}. However, the experimental characterization of such Frenkel-Wannier excitons has remained challenging, and much is unknown about the electronic composition and spatial characteristics of the interfacial excitonic wavefunctions. 

A highly promising platform for the realization of hybrid Frenkel-Wannier excitons is given by heterostructures of two-dimensional (2D) transition metal dichalcogenides (TMD) combined with organic semiconductors (OSC). Both these material classes are known for reduced electronic screening that leads to the formation of strongly Coulomb-bound excitons. In OSCs, predominantly Frenkel-type and charge-transfer excitons exist, which both derive their wavefunction from molecular orbitals and which are commonly restricted to only a single, or to just the neighboring molecules~\cite{Puschnig02prl, valencia23electronicstructure, Hummer2004}. Conversely, TMDs are known for hosting delocalized Wannier-type excitons whose wavefunctions are built up from Bloch states of the valence and conduction bands~\cite{Wang18rmp}. These individual properties make TMD/OSC heterostructures particularly promising for the realization of hybrid Frenkel-Wannier excitons, and it is predicted that both hybrid and charge transfer excitons can exist whose wavefunctions are composed of contributions of molecular orbitals of the OSC and valence/conduction band Bloch states of the TMD~\cite{Gonzalez22prm, Tanda23pssa}. However, the spatial structure of their wavefunction, \ie  whether the Frenkel- or Wannier-contributions to the wavefunction are more dominant, or if an exciton with both Frenkel and Wannier character can form and exist, remains largely unexplored. Moreover, experimental evidence for ultrafast charge- and energy-transfer processes is scarce.

Here, using femtosecond momentum microscopy~\cite{Keunecke20timeresolved} as well as \GW{} quasiparticle band structure~\cite{Hedin1965} and Bethe-Salpeter equation (BSE)~\cite{Rohlfing2000} calculations, we identify and characterize a hybrid Frenkel-Wannier exciton in the prototype system of monolayer 3,4,9,10-perylenetetracarboxylic dianhydride (PTCDA) adsorbed on monolayer tungsten diselenide (WSe$_2$).
We chose this system because, on the inorganic side, the electronic band structure~\cite{Wilson17sciadv}, the energy landscape of excitons~\cite{Wang18rmp}, and the resulting ultrafast exciton dynamics~\cite{Jin18natnano, Madeo20sci, Bange232DMaterials} are well characterized. Complementary, on the organic side, PTCDA is a key model system for the fabrication of flat molecular layers of organic semiconductors adsorbed on pristine surfaces~\cite{Tautz07pss} and for the study of optical excitations~\cite{Proehl04prl, Wallauer20sci}. This is an ideal setting to study the orbital contributions to all relevant optically bright- and dark-exciton wavefunctions. This includes the momentum-direct and -indirect intralayer excitons in WSe$_2$, and, intriguingly, the formation of a hybrid WSe$_2$/PTCDA exciton. Our joint experimental and theoretical results show that this hybrid exciton's wavefunction is a coherent superposition of intra- and interlayer contributions with Frenkel- and Wannier-character, respectively. Moreover, the orbital-resolved access to the exciton wavefunction combined with femtosecond time-resolution enables us to characterize the formation mechanism of the hybrid exciton. We show that in response to the optical excitation of WSe$_2$ A1s-excitons, exciton-phonon scattering and a Förster-type energy-transfer lead to the establishment of a steady state population between intralayer momentum-direct and -indirect WSe$_2$ excitons and the energetically most favorable hybrid exciton.

\vspace{.5cm}
\noindent\textbf{TMD/OSC sample structure and single-particle energy level alignment}
\vspace{.2cm}

\begin{figure*}[htbp]
    \centering
    \includegraphics[width=\textwidth]{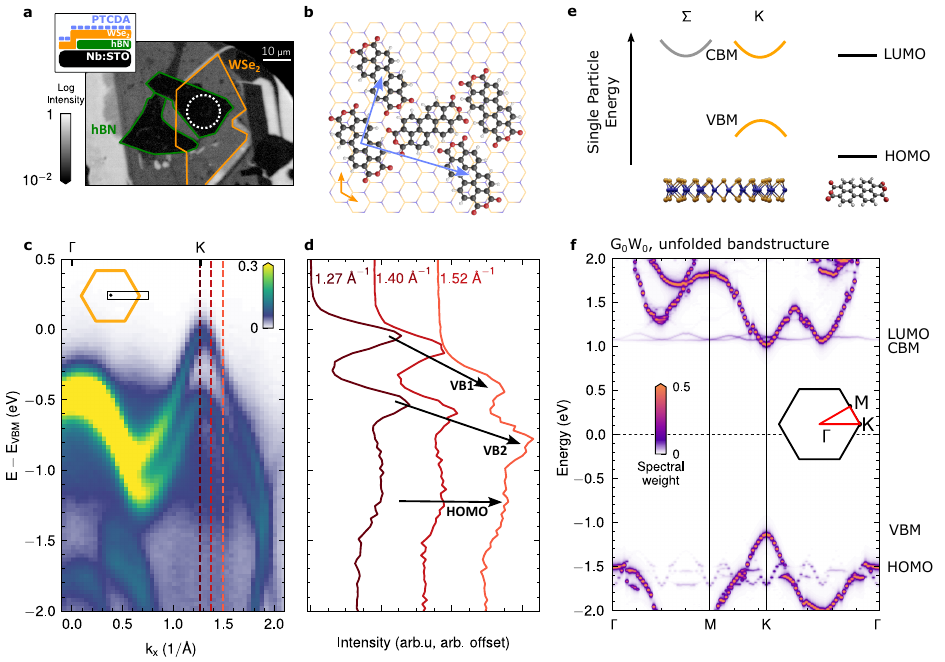}
    \caption{
    \textbf{Sample layout and electronic structure of the hybrid WSe$_2$/PTCDA heterostructure.}
    \textbf{a} Sketch of the layered sample structure and real-space photoemission image. The real-space region-of-interest addressed in the momentum-resolved photoemission measurement is marked by the white dashed circle, while the hBN flake and the WSe$_2$ monolayer are indicated by colored lines. 
    \textbf{b} Experimentally determined superstructure of PTCDA adsorbed on WSe$_2$ monolayer (cf. extended Fig.~\ref{extFig_SamplePrep}). 
    \textbf{c} Energy-momentum cut of the static photoemission spectrum along the $\Gamma$-K direction of the WSe$_2$/PTCDA heterostructure measured at 50 K. 
    \textbf{d} Energy distribution curves taken at the momenta indicated in \textbf{c}. The dispersive spin-split WSe$_2$ bands (VB1, VB2) and the non-dispersive HOMO level are marked by arrows. 
    \textbf{e} Overview of the type-I energy level alignment of the TMD/OSC heterostructure, as determined from static photoemission spectroscopy (\textbf{c,d}), the \GW{} calculation (\textbf{f}), and scanning tunneling spectroscopy experiments reported in refs.~\cite{Zheng16acsnano, Guo22nanoresearch}. 
    \textbf{f} Unfolded single-particle energy landscape of the WSe$_2$/PTCDA heterostructure as retrieved from the scissor-shifted \GW{} calculation in a 4$\times$4$\times$1 supercell (cf. extended Fig.~\ref{extFig_calcBandstructure}c and Methods). 
    }
\end{figure*}

The WSe$_2$ monolayer was fabricated by mechanical exfoliation and transferred on bulk hexagonal boron nitride (hBN) on a Nb:STO substrate (Fig.~1a, see Methods). Subsequently, close to a monolayer PTCDA was evaporated under ultra-high vacuum conditions. By analyzing the Umklapp scattering of the photoemitted electrons at the molecular superstructure (extended Fig.~1e), we verified that PTCDA adsorbs in an ordered Herringbone structure with a 17.94$\times$11.5~\r{A}{}$^2$ supercell (Fig.~1b). 

To extract the energy-level alignment of this WSe$_2$/PTCDA heterostructure, we start with static ARPES experiments (26.5~eV extreme ultraviolet (EUV) photons, 20~fs, $p$-polarized). The energy- and momentum-resolved photoemission data and the momentum-filtered energy distribution curves (EDCs) are shown in Figs.~1c,d. We find clear signatures of the K valley WSe$_2$ valence band maximum (VBM) (E-E$_{\rm VBM}=0$~eV) and additional spectral weight at an energy of E-E$_{\rm VBM}=-1.2 \pm 0.1$~eV that we identify with the highest occupied molecular orbital (HOMO) of PTCDA (see also momentum map in Extended Fig.~\ref{extFig_POT}f). 
We note that no signatures for hybridization of WSe$_2$ valence states and PTCDA orbitals are observed. The energy level alignment of the WSe$_2$ VBM and the PTCDA HOMO found here (Fig.~1e) is consistent with earlier scanning tunneling spectroscopy (STS) experiments~\cite{Zheng16acsnano, Guo22nanoresearch}, and, in addition, in reasonable agreement with our \GW{} calculations performed with the \exciting{} code~\cite{Gulans2014} (Fig.~1f and Methods). Combining all this information, we find that the single-particle energy-level alignment of the hybrid WSe$_2$/PTCDA heterostructure is of type I, where the energies of both the lowest unoccupied and the highest occupied state are found in WSe$_2$.

\vspace{.5cm}

\noindent\textbf{Momentum-resolved characterization of excitons at the 2D-organic interface}
\vspace{.2cm}

\begin{figure}[b]
    \centering
    \includegraphics[width=.5625\textwidth]{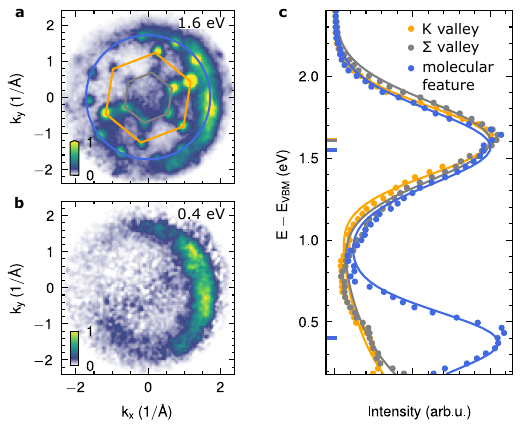}
    \caption{\textbf{Energy- and momentum-resolved identification of the excitonic photoemission signatures.} 
    \textbf{a},\textbf{b} Energy-filtered momentum maps and \textbf{c} momentum-filtered energy-distribution-curves of excitonic photoemission signatures. The data is obtained by integrating over the pump-probe delays from 100 to 500~fs and the application of the non-negative matrix factorization formalism (cf. Methods and extended Fig.~\ref{extFig_NMF}). 
    \textbf{a} The K and $\Sigma$ valleys of the WSe$_2$ Brillouin zone are indicated by the corners of an orange and a grey hexagon, respectively. The blue circle with a radius of $\sqrt{k_x^2+k_y^2}\approx 1.7$~\r{A}\textsuperscript{-1} corresponds the the expected mean radius of the simulated momentum distribution of the LUMO of PTCDA (cf. extended Fig.~\ref{extFig_POT}).
    \textbf{c} The EDCs are filtered in momentum for the K (orange), $\Sigma$ (grey), and molecular (blue) photoemission signatures (see extended Fig.~\ref{extFig_EDCs} for chosen region of interests) and fitted with a single or two Gaussian peaks (cf. Methods). The resulting peak energies are marked with a horizontal bar in the plot and the corresponding exciton energies E$_{\rm exc}^i$ are summarized in Table~\ref{table:fit_edcs}.
    }
\end{figure}

The extension of the static ARPES experiment with a femtosecond pump-probe scheme is used to characterize the orbital contributions to optical excitations at the 2D-organic interface. The experimental conditions are chosen such that the driving photon energy of 1.7~eV ($s$-polarized, 40~fs) lies well below the lowest-energy optical excitation in PTCDA, which is $\gtrapprox$2~eV~\cite{Proehl04prl}, \ie no PTCDA-only excitations can be created by the optical pump pulses. Instead, the laser pulses selectively excite the bright intralayer A1s-excitons in WSe$_2$, which we label as K-excitons here (peak fluence: 280$\pm$20 $\mu$Jcm$^{-2}$, exciton density: (5.4$\pm$1.0)$\times10^{12}$ cm$^{-2}$). All other excitons that are detected in the photoemission experiment must result from subsequent charge and energy transfer processes, as will be further discussed below. Figures~2a and 2b show time-delay-integrated photoemission momentum maps collected in selected energy windows above the WSe$_2$ valence bands. In the (k$_x$,k$_y$)-momentum-resolved data, we find a rich intensity structure that provides direct evidence for the presence of excitons that are of pure WSe$_2$ intralayer character, but also of excitons with distinct orbital contributions from the PTCDA layer.

First, we focus on the WSe$_2$ intralayer excitons. Momentum-sharp photoemission features are detected at the K and $\Sigma$ valleys of the WSe$_2$ Brillouin zone after optical excitation (Fig.~2a, corners of the orange and grey hexagons, respectively; $\Sigma$ also labeled as Q or $\Lambda$ in literature). The K valley spectral weight can be attributed to photoemitted electrons originating from optically bright K-excitons that we excite with the laser pulses (A1s-excitons), and momentum-indirect excitons where the electron- and hole-component reside at the K and K$^{\prime}$ valley, respectively. As photoemission spectral weight from these excitons both appear at the K (K$^{\prime}$) valley~\cite{Kunin23prl} and cannot be differentiated within the energy resolution of our experiment~\cite{Bange232DMaterials, Schmitt22nat}, we label those jointly as K-excitons  (Fig.~2d). Likewise, the $\Sigma$ valley spectral weight is indicative of the formation of momentum-indirect $\Sigma$-excitons whose electron- and hole-components reside in the $\Sigma$ and K valley, respectively~\cite{Madeo20sci, Wallauer21nanolett, Bange232DMaterials}. The energetic alignment of the K and $\Sigma$-excitons can be analyzed by evaluating momentum-filtered EDCs (Fig.~2c) and considering the conservation of energy as the EUV laser pulses fragment the excitons into their single-particle electron- and hole-components in the photoemission process~\cite{Reutzel24AdvPhysX} (E$_{\rm exc}^{\rm K}= 1.61\pm 0.05$~eV, E$_{\rm exc}^\Sigma=1.61\pm 0.05$~eV, cf. table \ref{table:fit_edcs} in Methods.).

Next, we turn our attention to the two semi-circular photoemission signatures at a radius of $\sqrt{k_x^2+k_y^2}\approx 1.7$~\r{A}\textsuperscript{-1} found in momentum maps with center energies of E-E$_{\rm VBM}=1.57\pm 0.05$~eV (Fig.~2a) and E-E$_{\rm VBM}=0.39\pm 0.05$~eV (Fig.~2b, corresponding EDCs in Fig.~2c). Such circular structures of spectral weight are characteristic for photoelectrons emitted from molecular orbitals~\cite{Puschnig09sci, Wallauer20sci} (cf. extended Fig.~\ref{extFig_POT}c). Notably, the features are not found in the case of pristine monolayer WSe$_2$, but only in the case of the WSe$_2$/PTCDA heterostructure (extended Fig.~\ref{extFig_POT} and refs.~\cite{Madeo20sci, Bange232DMaterials}). Since the excitation energy of 1.7~eV is well below the direct HOMO$\rightarrow$LUMO excitation ($\gtrapprox$2~eV~\cite{Proehl04prl}), these PTCDA orbital-like photoemission signatures are expected to result from a charge- or energy-transfer process across the TMD/OSC interface, and, in consequence, are of major interest to our study.

\vspace{.5cm}
\noindent\textbf{A hybrid exciton bridging the 2D-organic interface}
\vspace{0.2cm}

The question at hand is in how far these two fingerprints of molecular orbitals are an indication for multiple excitons with either interlayer or pure PTCDA character, or if they are a fingerprint of a so far only predicted hybrid Wannier-Frenkel excitons~\cite{Agranovich11cr} with multiple hole contributions~\cite{Bennecke24natcom, Kern23prb, Meneghini23ACSPhotonics, Neef23nat} from WSe$_2$ and PTCDA.
To address this question, we start by solving the Bethe-Salpeter equation on top of our \GW{} calculations of the WSe$_2$/PTCDA heterostructure in a 4$\times$4$\times$1 supercell (extended Fig.~\ref{extFig_calcBandstructure}a, cf. Methods). These calculations yield the imaginary part of the in-plane frequency-dependent dielectric function Im($\epsilon_{||}\left(q\right)$), containing information on optical excitations at the 2D-organic interface, and allows us to identify the band/orbital contributions to the exciton wavefunctions in momentum- and real-space~\cite{Gonzalez22prm}. In the momentum-direct part of the spectrum (\ie with $q = k_e-k_h = 0$), the two lowest-energy excitons are of WSe$_2$ (orange, $E_{\rm exc}^{\rm K,BSE}=1.74$~eV) and hybrid (blue, $E_{\rm exc}^{\rm hX,BSE}=1.72$~eV) character, respectively (Fig.~3a, marked with arrows). In Fig.~3c,d the electron and hole contributions to these excitons are analyzed in reciprocal space. We find that the 1.74~eV K-exciton is of full WSe$_2$ intralayer character and derives its wavefunction purely from WSe$_2$ conduction and valence band Bloch states (Fig.~3c). In contrast, for the 1.72~eV exciton, which we term hybrid exciton (hX), the electron component derives its wavefunction from the LUMO of PTCDA, while the hole component has contributions from the WSe$_2$ valence bands and also from the PTCDA HOMO (Fig.~3d). Hence, the \GW{}+BSE calculations imply that both intralayer HOMO$\rightarrow$LUMO and interlayer VBM$\rightarrow$LUMO transitions are mixed into an energetically favorable hX with contributions from both sides of the interface. 
We note that, at first glance, such a mixing might seem counter-intuitive because of the large energy difference of the intra- and interlayer single-particle band gaps (\ie ${\rm E_{LUMO}-E_{HOMO}>E_{LUMO}-E_{VBM}}$, Fig.~3e). However, due to the stronger electron-hole interaction for the case of intralayer HOMO$\rightarrow$LUMO transitions as compared to interlayer VBM$\rightarrow$LUMO transitions, the individual exciton energies of both electron-hole transitions can be sufficiently degenerate to allow the mixing of both OSC and TMD components (Fig.~3f).

\begin{figure}
    \centering
    \includegraphics[width=.99\textwidth]{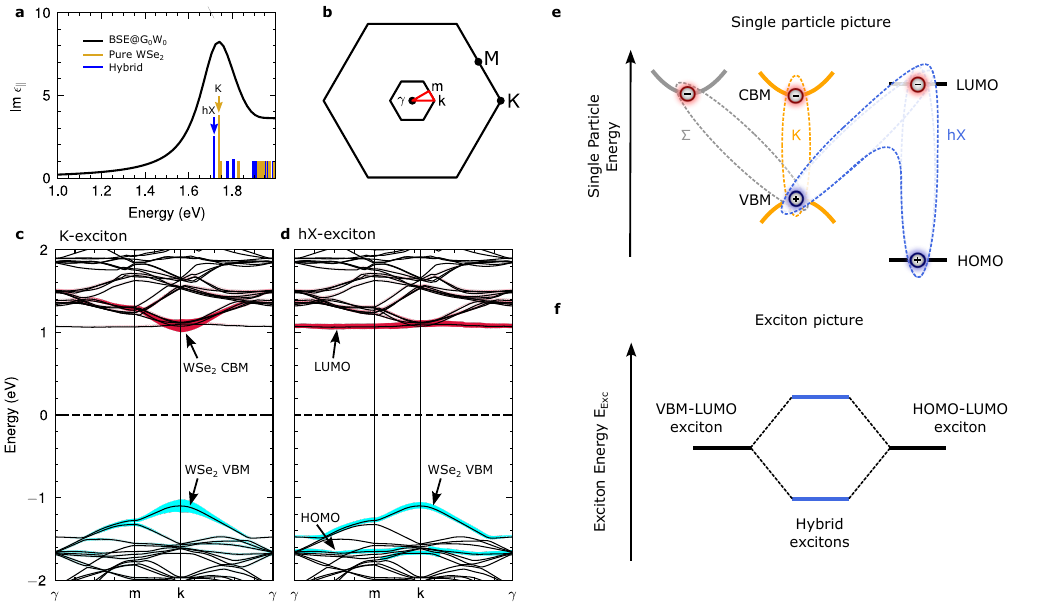}
    \caption{\textbf{Reciprocal-space representation of the Bloch states and molecular orbitals contributing to the K-exciton and the hX wavefunction.}
    \textbf{a} Absorption spectrum of WSe$_2$/PTCDA retrieved by G$_0$W$_0$+BSE calculations. The oscillator strengths of the contributing excitons are indicated as solid lines where all values below one (dark-excitons) are set to one for visibility. Excitons with and without contributions from PTCDA orbitals are distinguished in blue and yellow, respectively. 
    \textbf{b} Backfolded Brillouin zone according to the theoretical superstructure (see extended Fig.~\ref{extFig_calcBandstructure}a).
    \textbf{c,d} The two lowest-lying excitons marked by arrows in \textbf{a} are analyzed in detail in reciprocal space using the backfolded Brillouin zone (\textbf{b}). The electron and hole contributions are marked in red and cyan, respectively. \textbf{c} While the K-exciton wavefunction is purely composed of TMD valence and conduction band states (WSe$_2$ VBM and CBM), \textbf{d} the hX wavefunction has contributions from the TMD valence bands (WSe$_2$ VBM) and from the PTCDA HOMO and LUMO orbitals. 
    \textbf{e} Visualization of the electron-hole transitions that contribute to the wavefunction of K-exciton, $\Sigma$-exciton, and hybrid exciton (hX). The hX wavefunction is of partial intra- and interlayer composition and built up by HOMO$\rightarrow$LUMO and VBM$\rightarrow$LUMO transitions, respectively. 
    \textbf{f} Illustration of the hX in the exciton picture. The intralayer and interlayer electron-hole transitions are expected to be nearly degenerate in energy because of the stronger electron-hole interaction of the pure HOMO-LUMO exciton compared to the VBM-LUMO exciton. Mixing of the transitions leads to the formation of a new bound hybrid excitonic state at lower exciton energies, \ie the hX.
    }
\end{figure}

\vspace{.5cm}
\noindent\textbf{Experimental characterization of the hX}
\vspace{0.2cm}

The \GW{}+BSE prediction of the dual-component character of the hX is in excellent agreement with our experimental findings, as can be verified by analyzing three characteristic photoemission fingerprints in the (i) momentum-, (ii) energy-, and (iii) time-delay-domain. First (i), the measured momentum maps shown in Fig.~2 are both in agreement with the LUMO orbital momentum map calculated within the framework of orbital tomography~\cite{Puschnig09sci} (see also extended Fig.~\ref{extFig_POT}), clearly confirming experimentally that the exciton's electron component resides in the molecular layer~\cite{Kern23prb}. Second (ii), it is known that photoelectrons that are emitted from excitons 
are detected one exciton energy E$_{\rm exc}^i$ above the energy of the single-particle bands where the hole component remains after photo excitation~\cite{Kern23prb, Bennecke24natcom, Meneghini23ACSPhotonics}. Thus, for the hX, which has two hole contributions from the WSe$_2$ VBM and the PTCDA HOMO, one can expect to observe not only one, but two photoemission orbital signatures that are separated in energy by the energy difference between the WSe$_2$ VBM  and the PTCDA HOMO level: $\Delta E_{\rm VBM,HOMO}=E_{\rm VBM}-E_{\rm HOMO}=1.2\pm 0.1$~eV. Strikingly, the experimentally found peak-to-peak energy difference $\Delta E_{\rm hX}=1.18\pm 0.08$~eV (Fig.~2a) of the two excitonic photoemission signatures is in quantitative agreement with $\Delta E_{\rm VBM,HOMO}$. Third (iii), if both photoemission signatures result from the break-up of the same hybrid exciton, their population dynamics have to coincide. Indeed, as we will discuss later in detail, the delayed onset and decay dynamics of both photoemission signatures are in quantitative agreement (cf. Fig.~5, extended Fig.~\ref{extFig_tempAnalysis}, Methods Table~\ref{table:fit_dynamics}). We therefore conclude from experiment and theory that the WSe$_2$/PTCDA heterostructure hosts a hybrid exciton whose wavefunction extends across the TMD/OSC interface.

\begin{figure*}
    \centering
    \includegraphics[width=.99\linewidth]{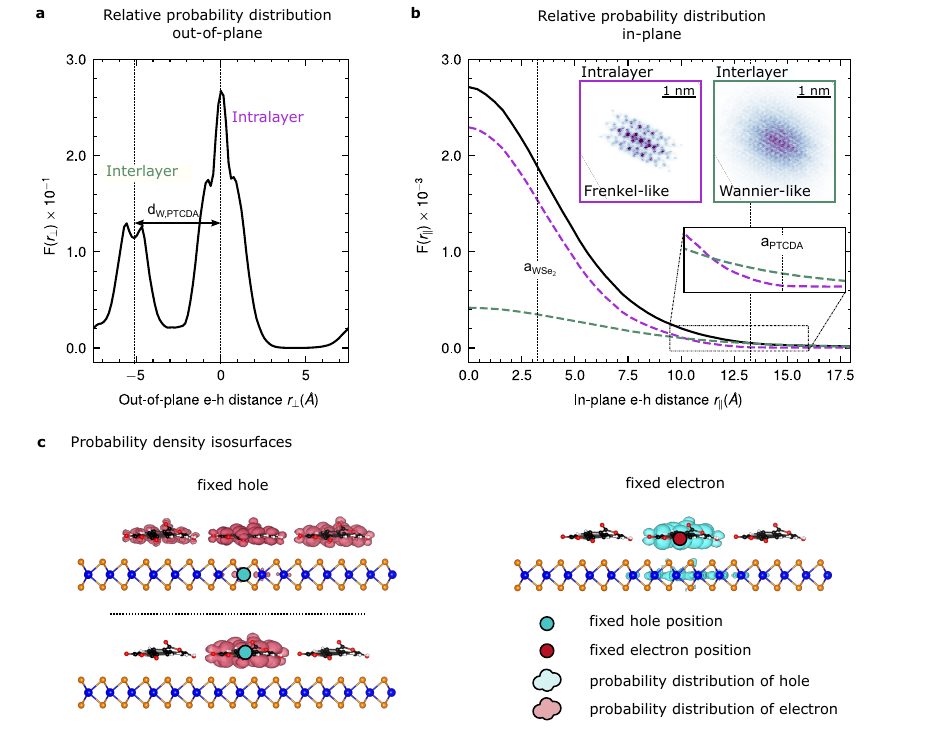}
    \centering
    \caption{
    \textbf{Real-space properties of the hX wavefunction}.
    \textbf{a} The out-of-plane component of the electron-hole correlation function $F^i(\ve{r}) = F(\ve{r}_e-\ve{r}_h)$ shows two peaks separated by the distance between the tungsten plane and the PTCDA molecule (d$_{\rm W,PTCDA}$), which confirms a combination of both intralayer ($r_\perp\approx 0$~\r{A}) and interlayer ($r_\perp\approx -5$~\r{A}) character.  
    \textbf{b} By splitting the in-plane electron-hole correlation into intra- (purple) and interlayer (green) contributions it is possible to visualize the the major difference in spatial extend. 
    To compare these spatial distributions with the underlying atomic structure, the lattice constants of WSe$_2$ and the PTCDA superstructure are indicated as dashed vertical lines. The insets show the $r_\parallel=$(r$_x$,r$_y$) resolved representation of the Wannier-type (green axis) and the Frenkel-type (purple axis) contributions.
    \textbf{c} Exemplary probability density isosurfaces for the hX for fixed hole (left) and fixed electron (right) position. Due to the dual Wannier-Frenkel character, the hX probability density isosurfaces depend strongly on the chosen hole location, where the isosurface extends over multiple molecules when the hole is placed in a TMD Bloch state (top), while the isosurface shows clear Frenkel nature when the hole is placed in the PTCDA HOMO (bottom). If the electron is fixed at the molecule (right), the isosurface of the hole has contributions both in the TMD and the PTCDA molecule.
    }
\end{figure*}

\vspace{.5cm}
\noindent\textbf{Real-space wavefunction distribution of hybrid Wannier-Frenkel excitons}
\vspace{0.2cm}

Having access to the excitonic wavefunction contributions of the WSe$_2$ Bloch states and the PTCDA orbitals from experiment and theory, we are in the position to evaluate the real-space Frenkel- and/or Wannier-character of the hX. Specifically, we aim to characterize the exciton's relative electron-hole distance parallel and perpendicular to the heterostructure in comparison to the size of the WSe$_2$ and PTCDA unit cells. 
Therefore, we analyze the electron-hole correlation function, \ie the probability distribution of the electron-hole separation, $F^i(\ve{r}) = F(\ve{r}_e-\ve{r}_h)$ with regard to the heterostructure's out-of-plane ($r_{\perp}$, Fig.~4a) and in-plane ($\ve{r}_{\parallel}$, Fig.~4b) coordinates~\cite{Sharifzadeh13jpcl}, which correspond to the excitons intra-/interlayer and Frenkel/Wannier character, respectively (details in Methods; analysis for K-exciton in Extended Fig.~\ref{extFig_Aexciton_spatial}). 

Intriguingly, for the out-of-plane component (Fig.~4a), there is not only a peak around $r_\perp\approx0$~\r{A} that indicates intralayer character, but also a peak centered at $r_\perp\approx -5$~\r{A}, which matches the distance between the tungsten plane and the PTCDA molecule and therefore indicates the additional interlayer character of the hX. Hence, the double-peak structure in Fig.~4a is a direct signature of the mixed intra- and interlayer contributions to the hX. 

Complementary, the in-plane electron-hole distribution function $F^i(\ve{r}_{\parallel})$ (Fig.~4b) contains information on the Frenkel- and/or Wannier character of the hX. Here, we plot the intra- and interlayer contributions separately as purple and green lines in Fig.~4b. While the relative probability of the intralayer contribution is completely confined to values smaller than the lattice constant of the PTCDA supercell (a$_{\rm PTCDA}$) indicating a Frenkel-like character, the interlayer contribution is more delocalized and indicates a Wannier-like character (Fig.~4b and zoom-in around a$_{\rm PTCDA}$). The different character is even more evident when considering the two-dimensional representations of $F^{\rm hX}(\ve{r}_{\parallel})$ in the insets in Fig.~4b, which show that the interlayer-component resembles the overall Gaussian intensity distribution known for a K-exciton (cf. extended Fig.~7b, refs.~\cite{Man21sciadv, dong20naturalsciences, Schmitt22nat}), while the Frenkel-component shows a much stronger spatial structuring that stems from the molecular orbital. 

For further illustration of the spatial extension of the hX, both in terms of intra- and interlayer contributions, and also its Wannier and Frenkel nature, Fig.~4c shows exemplary probability density isosurfaces of the exciton's electron- and hole-components (red shaded and cyan shaded volumes) that are obtained by fixing the exciton's hole (cyan dots) or electron components (red dot) at typical positions in the heterostructure. When the hole is placed in a delocalized state in the TMD layer (top left), we find a Wannier-like isosurface where the electron is spread over multiple PTCDA molecules. In contrast, when the hole is placed on the PTCDA molecule (bottom left), the electron is completely localized on the same molecule, too. In other words, the isosurface now describes a Frenkel-type exciton. Most interestingly, when the electron is fixed on the PTCDA layer (right), the hole isosurface again shows the dual component characteristics with a localized part on the same PTCDA molecule, and also a (weaker) delocalized contribution on the TMD layer. Hence, the hole isosurface has both Wannier and Frenkel character. This hybrid intra- and interlayer and Frenkel- and Wannier-nature of the hX is a unique feature of the TMD/OSC interface that highlights the versatility of these combined platforms for controlling optoelectronic energy conversion pathways.

\vspace{1cm}

\noindent\textbf{Femtosecond time- and orbital-resolved exciton dynamics}
\vspace{0.2cm}

\begin{figure}[b]
    \centering
    \includegraphics[width=.5625\textwidth]{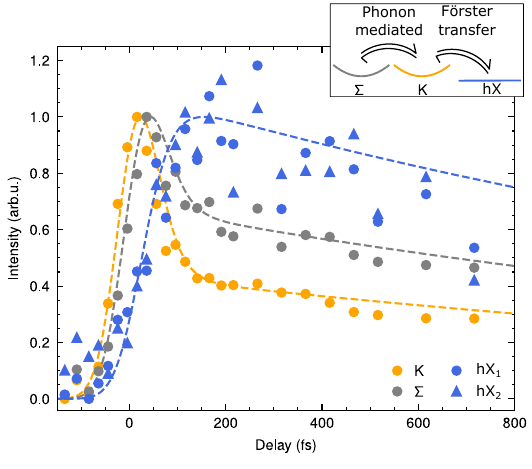}
    \caption{
    \textbf{Femtosecond formation of the hX at the hybrid WSe$_2$/PTCDA interface}.
    Subsequent to the optical excitation of WSe$_2$ K-excitons (orange), exciton-phonon scattering and Förster-type dipole-dipole interactions lead to the formation of $\Sigma$-excitons (grey) and the hX (blue), respectively. The symbols encode photoemission spectral weight filtered in energy- and momentum space (cf. Extended Fig.~\ref{extFig_EDCs}), and the dashed lines are guides to the eye. A steady state between all excitonic occupations is reached after 150~fs (inset), which decays on the 2~ps timescale. The formation and decay rates of all excitonic photoemission signatures are quantified in extended Fig.~\ref{extFig_tempAnalysis} and summarized in Table~\ref{table:fit_dynamics}. 
    }
\end{figure}

Finally, we want to elucidate the ultrafast formation and thermalization dynamics of all excitons involved. Figure~5 shows the femtosecond pump-probe delay dependence of photoemission spectral weight from K-excitons (K valley), $\Sigma$-excitons ($\Sigma$ valley), and hXs (molecular features hX$_1$ and hX$_2$). Subsequent to the optical excitation of K-excitons and the concomitant rise of spectral weight at the K valley (orange), the photoemission intensities at the $\Sigma$ valley (grey) and for the molecular features (blue) rise with delayed onsets of 19$\pm$4~fs and 66$\pm$11~fs, respectively. The molecular features' spectral weight peaks at about 150~fs, and their decay can be well-described with a single-exponential function with a decay constant of $\tau_{\rm decay}^{\rm hX}=1.9\pm 0.7$~ps. Interestingly, following the increase of the hX spectral weight, the initially fast decay of the K and $\Sigma$ valley spectral weight is slowed down and shows the same behavior as the decay of the hX on longer timescales. This is confirmed by fitting bi-exponential functions to the decay of the K- and the $\Sigma$-exciton, which show slow decay constants of $\tau_{\rm slow}^{\rm K}=2.1\pm 0.4$~ps and $\tau_{\rm slow}^{\rm \Sigma}=2.2\pm 0.4$~ps, respectively (Methods Table \ref{table:fit_dynamics}). From this analysis, we conclude that subsequent to the optical excitation of K-excitons and on a sub-200~fs timescale, a K$\rightleftharpoons\Sigma\rightleftharpoons$hX steady state population with common decay channels is established.

However, while in pure TMD heterostructures charge and energy transfer across interfaces is mediated and explained by band structure hybridization effects~\cite{Schmitt22nat, Bange24SciAdv, Meneghini22naturalsciences, Wang17prb}, the general preconception for TMD/OSC interfaces is that orbital hybridization is weak~\cite{Krumland21electronicstructure, Park21AdvSci}, and that, in consequence, related charge-transfer processes are inhibited. Indeed, neither in experiment nor in theory, we find signatures of orbital hybridization of TMD/OSC excitons: The experimentally determined K- and $\Sigma$-exciton energies are in quantitative agreement with ARPES~\cite{Bange232DMaterials} and photoluminescence~\cite{Li14prb} data reported for monolayer WSe$_2$ (Table~~\ref{table:fit_edcs}), \ie without a molecular overlayer. Specifically, compared to E$_{\rm exc}^K$, no reduction of E$_{\rm exc}^\Sigma$ due to the PTCDA adsorption is observed, as is the case of pure TMD few-layer systems with hybridized bands~\cite{Bange232DMaterials}.
Hence, the experimental results imply that the K- and $\Sigma$-exciton wavefunctions are of mere WSe$_2$ intralayer character, with negligible contribution from PTCDA orbitals. This is in agreement with our DFT calculations that also do not show any hybridization of the single particle PTCDA LUMO and WSe$_2$ conduction band states (extended Fig.~\ref{extFig_calcBandstructure}c).
In consequence, the formation of the hX cannot be mediated by hybridized states between the layers. Instead, we conclude that energy transfer must be mediated by dipole-dipole interactions, \ie in a Förster-type energy transfer process (FRET). Indeed, all requirements for such a process are fulfilled~\cite{Katzer23prb, Thompson23npj2d}: First, the K- and hX- energies are close in energy, satisfying the requirement of energy conservation (Fig.~2c and Methods Table~\ref{table:fit_edcs}). Second, the dipole moment of the HOMO-LUMO contribution to the hX is polarized in-plane, facilitating the coupling to the in-plane dipole-moment of the WSe$_2$ K-exciton. And third, because of the mixed nature of the hX wavefunction and the contribution of HOMO-LUMO transitions, our \GW{}+BSE calculations predict that the oscillator strength is rather large (Fig.~3a), making the Förster-type process efficient. We note, however, that the hX formation is 1-2 orders of magnitude faster than typically found in other reports discussing the dynamics of Förster processes at TMD/OSC~\cite{Katzer23prb} or TMD/graphene~\cite{Selig19prb, Dong23natcom} interfaces. We suspect that the fast rise time of the hX might be related to the fact that the dipole-dipole interactions do not induce a complete energy-transfer across the 2D-organic interface. Instead, the dipole-dipole interactions promote the conversion of WSe$_2$ intralayer K-excitons to hXs, whose wavefunction is composed of PTCDA intralayer HOMO/LUMO and interlayer WSe$_2$-VBM/LUMO orbital contributions. 
While future theoretical work is needed to verify the timescale of the Förster-type energy-transfer process at the WSe$_2$/PTCDA interface, our work already highlights how excitonic wavefunction engineering can directly contribute to efficient energy transfer processes in 2D/organic hybrid heterostructures.

\clearpage

\vspace{1cm}
\section*{METHODS}
\vspace{1.cm}

\noindent\textbf{Sample preparation}
\vspace{0.2cm}

To fabricate the WSe$_2$/PTCDA heterostructure (extended Fig.~1a), hBN was first exfoliated onto a 0.1\% niobium-doped SrTiO$_3$ substrate and a $\approx$50~nm thick flake was identified by optical microscopy. In a parallel procedure, WSe$_2$ monolayers were directly exfoliated onto a silicone gel-film (DGL Film, Gel-Pak) and identified through optical microscopy and Raman spectroscopy. Afterwards, a monolayer WSe$_2$ flake was transferred from the silicone gel-film onto the hBN flake on the SrTiO$_3$ substrate (extended Fig.~1b). After introduction into ultra-high vacuum (<~$5\cdot 10^{-9}$~mbar), the sample was annealed at 670~K for 2~hours to ensure a clean sample surface. The bare monolayer WSe$_2$ was analyzed with the momentum microscope in real (extended Fig.~\ref{extFig_SamplePrep}c) and reciprocal space (extended Fig.~\ref{extFig_POT}e), showing the expected characteristic features of monolayer WSe$_2$, \ie the spin-split valence bands at the K valley and a single parabolic band at the $\Gamma$ valley below the global valence band maximum at the K valley (cf. Fig.~1b and refs.~\cite{Bange232DMaterials, Wilson17sciadv}).  

Subsequently, close to a monolayer PTCDA was thermally evaporated onto the sample, which was kept at room temperature (base pressure <~$1\cdot 10^{-9}$~mbar). The deposition rate was monitored with a with a quartz crystal microbalance and calibrated using the known deposition of PTCDA onto a Ag(110) crystal surface. On the Ag(110) surface, the first monolayer of PTCDA is adsorbed in a brickwall structure whereas additional layers grow in a Herringbone structure with a different superstructure that can be analyzed by low energy electron diffraction (LEED)~\cite{WiessnerPRB2012}. By step-wise evaporation of PTCDA onto Ag(110) and recording of the LEED pattern, the evaporation rate was determined, and was then used to deposit a monolayer PTCDA onto monolayer WSe$_2$. The successful deposition of a monolayer PTCDA onto monolayer WSe$_2$ was confirmed by the observation of additional spectral weight in the static ARPES data, which can be attributed to the HOMO of PTCDA (Fig.~1c,d) and backfolded WSe$_2$ bands (extended Fig.~\ref{extFig_SamplePrep}e), which are caused by the adsorbed molecular PTCDA superstructure (Fig.~1b). The superstructure matrix 
\begin{align*}
M = \begin{bmatrix}
    1.58       & 6.78 \\
    4.39       & 1.32 
    \end{bmatrix}    
\end{align*}
was determined by LEED on a cleaved WSe$_2$ bulk crystal (extended Fig.~\ref{extFig_SamplePrep}d) and is in good agreement with ref.~\cite{ZhangNanoResearch2022} and the ARPES data (extended Fig.~\ref{extFig_SamplePrep}e). 

\vspace{1cm}
\noindent\textbf{Femtosecond momentum microscopy}
\vspace{0.2cm}

All photoemission data were acquired with the Göttingen in-house photoemission setup~\cite{Keunecke20timeresolved,Keunecke20prb} that combines a time-of-flight momentum microscope~\cite{medjanik_direct_2017} (Surface Concept GmBH, ToF-MM) with a 500~kHz high-harmonic generation beamline (26.5~eV $p$-polarized, 20~fs). For the time-resolved measurements, the photon energy of the $s$-polarized pump was tuned to h$\nu$ = 1.7~eV with 50$\pm$5~fs pulse duration using an optical parametric amplifier. The pump fluence was adjusted to 280$\pm$20 $\micro$J cm$^{-2}$, which results approximately in an initial K-exciton density of $(5.4 \pm 1.0) \times 10^{12}$~cm$^{-2}$~\cite{Li14prb, Schmitt22nat}. The static measurements (Fig.~1c) were performed at $T = 50$~K, while all pump-probe delay-dependent measurements were performed at room temperature (300~K). 

\vspace{1cm}
\noindent\textbf{Photoemission data processing}
\vspace{0.2cm}

The time-of-flight momentum microscope enables the simultaneous measurement of the kinetic energy and both in-plane momenta of the photoemitted electrons~\cite{medjanik_direct_2017}. However, the acquired three-dimensional photoemission data are subject to several lens aberrations and other distortions such as pump and probe induced space-charge and surface photovoltage effects~\cite{schonhense_space-_2015, Roth24prl, Schmitt23arXiv}. Therefore, the photoemission data needs to be preprocessed before further evaluation by (i) correcting a time-dependent rigid energy shift and (ii) correcting for distortions that are induced by the projection and focal lens system.  

First (i), the time-dependent energy shift was corrected by minimizing the variance between the momentum-integrated spectra for E - E$_{\rm VBM}$ < 1.8 eV. Second (ii), an additional measurement was performed with a grid inserted in the Fourier plane~\cite{Maklar20rsi,Karni22nat}. We then determined the parameters for an affine transformation that maps the measured data onto an undistorted and energy-independent grid. This transformation is applied to all data sets. Small remaining distortions induced by the first lens system were corrected by fitting the positions of the K-excitons and mapping them onto an equilateral hexagon. The same positions were used to perform the momentum calibration using the lattice constant of WSe$_2$ $a_{\rm WSe_2}=0.3297$~nm. In addition, for each delay step, the data were momentum-wise normalized to the energy range between E-E$_{\rm VBM}$= $- 1.8$ and $-3$~eV. This momentum-wise normalization accounts for potential changes in illumination due to possible instabilities during the long integration times of the time-resolved measurements.

\vspace{1cm}
\noindent\textbf{Quantitative analysis of the exciton energies and dynamics}
\vspace{0.2cm}

The EUV laser pulses fragment the Coulomb-bound electron-hole pairs into their single-particle components. As this process conserves energy and momentum~\cite{Weinelt04prl,Rustagi18prb,Reutzel24AdvPhysX}, the exciton energies E$_{\rm exc}$ can be extracted by fitting the delay-integrated (100-500~fs), background substracted (see NMF method below), and momentum-filtered EDCs shown in Fig.~2c with either one (K- and $\Sigma$-excitons) or two (hX) Gaussian peaks $I_{\rm p}$ and an exponential background $I_{\rm bg}$, \ie

\begin{align}
    I_{\rm p} (E)  &= \frac{A}{\sigma\sqrt{2\pi}} \exp{\left(-\frac{(E-\mu)^2}{2\sigma^2}\right)}, \\
    I_{\rm bg} (E) &= A_{\rm bg}\exp{\left(-\frac{-E}{\tau_{\rm bg}}\right)}.
    \label{eq:EDC_fit}
\end{align}

The extracted peak energies E-E$_{\rm VBM}$ of the K and $\Sigma$ excitons directly correspond to the exciton energies E$_{\rm exc}$ since the hole resides at the VBM of WSe$_2$. For the hX, the the peak energy of the higher lying peak at E-E$_{\rm VBM}$ = 1.57 $\pm$ 0.05 eV directly corresponds to E$^{\rm hX}_{\rm exc}$, whereas the lower energy peak at E-E$_{\rm VBM}$ = 0.35  $\pm$ 0.05 eV has to be referenced to the HOMO at E-E$_{\rm VBM}$ = -1.2 $\pm$ 0.1 eV, which results in the same exciton energy E$^{\rm hX}_{\rm exc}$ = 1.55 $\pm$ 0.1 eV. In Table~\ref{table:fit_edcs}, the quantified exciton energies E$^i_{\rm exc}$are compared to the BSE@\GW{} calculations and ARPES and photoluminesence (PL) experiments on monolayer WSe$_2$~\cite{Bange232DMaterials, Madeo20sci, Karni19prl}. The total error of the experimental values is estimated to be approximately 0.05~eV, taking into account fitting errors and possible errors induced by the energy calibration and space charge effects. 

\begin{table}[h]
\caption{Comparison of the experimental and theoretical exciton energies E$^i_{\rm exc}$ of WSe$_2$/PTCDA to monolayer WSe$_2$ extracted from refs.~\cite{Bange232DMaterials, Madeo20sci, Karni19prl}. 
 }
\begin{tabular}{c>{\centering}m{2.5cm}>{\centering}m{2.5cm}>{\centering}m{2.5cm}>{\centering}m{2.5cm}c}
& \multicolumn{2}{c}{WSe$_2$/PTCDA} & \multicolumn{3}{c}{Monolayer WSe$_2$}\\
  & Experiment & Theory &  trARPES~\cite{Bange232DMaterials} & trARPES~\cite{Madeo20sci} & PL~\cite{Karni19prl}  \\ \hline \hline
 E$^{\rm K}_{\rm exc}$   (eV)   & $1.61 \pm 0.05$ & 1.74 & 1.67 $\pm$ 0.05& 1.73 & 1.66\\
 E$^\Sigma_{\rm exc}$   (eV)    & $1.61 \pm 0.05$ & $-$ & 1.60 $\pm$ 0.05& 1.73 &\\
 E$^{\rm hX}_{\rm exc}$ (eV)  &  $1.57 \pm 0.05$ & 1.72& & \\
\end{tabular}
\label{table:fit_edcs}
\end{table}
\clearpage

To analyze the exciton dynamics of the K, $\Sigma$ and hX photoemission signatures (cf. Fig.~5), we filter the raw photoemission data, \ie without background subtraction, by their energy and momentum coordinate. The respective energy distribution curves (EDCs) and the chosen region of interests are shown in extended Fig.~\ref{extFig_EDCs}. To quantify the rise time, we fitted the energy- and momentum-filtered time-resolved photoemission spectral weight traces with an error function
\begin{align}
    I(t) = \frac{1}{2}\left(1+\text{erf}\left(\frac{t-\mu_{\rm onset}}{\sqrt{2}\sigma_{\rm rise}}\right)\right)
\end{align}
in the delay regions -200~fs to 0~fs, -200~fs to 20~fs, and -200~fs to 150~fs, respectively (extended Fig.~\ref{extFig_tempAnalysis}). Here, $\mu_{\rm onset}$ indicates the onset time, while $\sigma_{\rm rise}$ is directly related to the rise time.

Similarly, we fitted the decay of photoemission spectral weight with a biexponential decay (eq. \ref{eq:biexp}) between 0~fs to 2000~fs and 20~fs to 2000~fs for the $K$ and $\Sigma$-excitons, respectively. The hX was fitted with a single exponential decay function (eq. \ref{eq:exp}) between 200~fs to 2000~fs:
\begin{align}
    I(t) &= A\left(\frac{1}{1+f}\exp{\left(-\frac{t}{\tau_{\rm fast}}\right)} + \frac{f}{1+f}{\left(-\frac{x}{\tau_{\rm slow}}\right)}\right) \label{eq:biexp},\\ 
    I(t) &= A\exp{\left(-\frac{t}{\tau_{\rm slow}}\right)}. \label{eq:exp}
\end{align}
The relevant time constants are given in table \ref{table:fit_dynamics}. 

\begin{table*}[h]
\caption{Fit parameters from the (bi)exponential decay and the error function rise time fits of the exciton dynamics shown in Extended Fig.~\ref{extFig_tempAnalysis}. For comparison, the fit parameters for the pure WSe$_2$ monolayer were taken from~\cite{Bange232DMaterials} and are also listed in the table.}
\begin{tabular}{lc >{\centering}m{2cm} >{\centering}m{2cm} >{\centering}m{2cm} c}
\multicolumn{2}{c}{} 
& $\tau_\text{fast}$ (fs) & $\tau_\text{slow}$ (ps) & $\mu_\text{onset}$ (fs) & $\sigma_\text{rise}$ (fs)\\ \hline \hline
\multirow{3}{*}[2em]{WSe$_2$/PTCDA} 
& $K$       &  $51 \pm 6$       & $2.1 \pm 0.4$     & $ -38 \pm 2$  & $20 \pm 3$ \\  
& $\Sigma$  &  $60 \pm 20$      & $2.2 \pm 0.4$     & $-19 \pm 3$   & $30 \pm 5$ \\  
& $hX_1$    &                   & $1.8 \pm 0.5$     & $28 \pm 6$    & $68 \pm  8$\\  
& $hX_2$    &                   &  $1.9 \pm 0.5$    & $27 \pm 7$    & $70 \pm 10$\\  \hline 
\multirow{3}{*}[2em]{Monolayer WSe$_2$} 
& $K$       &  $70 \pm 10$      & $1.5 \pm 0.2$     & $ -33 \pm 3$  & $23 \pm 3$ \\  
& $\Sigma$  &                   & $1.05 \pm 0.06$   & $-7 \pm 2$    & $33 \pm 3$ \\ 
\end{tabular}
\label{table:fit_dynamics}
\end{table*}

\clearpage
\vspace{1cm}
\noindent\textbf{Non-negative matrix factorization}
\vspace{0.2cm}

Due to the small size of the WSe$_2$ monolayer and the small real-space selection aperture (10~$\mu$m effective diameter), the measurement was susceptible to a time-independent background intensity. Therefore, the excited state momentum maps and momentum-filtered EDCs shown in Fig.~2 and  in the insets of extended Fig.~\ref{extFig_EDCs} and \ref{extFig_tempAnalysis} were background subtracted.

For the background determination, we used the non-negative matrix formalism (NMF), as implemented in the scikit-learn package for Python~\cite{fevotte2011NMF}. NMF is a dimensionality-reduction method that so far remains unexplored in time-resolved ARPES, but it has recently found application in spatially-resolved material science studies based on x-ray diffraction~\cite{kutsukake_feature_2024} and also static ARPES experiments~\cite{imamura_unsupervised_2024}. Similar to principal component analysis, NMF is based on the numerical factorization of a given matrix $X$ into two matrices $W$ and $H$, with the additional condition that all matrices have only non-negative elements. In our case, $X$ is given by the time-dependent raw data set where we only consider excited state data above E-E$_{\rm VBM}$ =  0.15~eV. Additionally, we fix $W$ by a static background and the four extracted time traces of the $K$, $\Sigma$, hX$_1$, and hX$_2$ photoemission signal plotted in Fig.~5. The determined output  then is the matrix $H$ that consists of five components, each following one of the four given time dependencies and the time-independent background. Extended Fig.~\ref{extFig_NMF} shows extracted components integrated over the regions of interest in energy. Notably, the different components 1-4 can be assigned in reasonable agreement to the different excitonic photoemission signatures despite the strong overlap in time, energy, and momentum (cf. orange/grey hexagons and blue circle). Component 5 is time-indepent and used for background substraction.

\vspace{1cm}
\noindent\textbf{Photoemission orbital tomography}
\vspace{0.2cm}

Using the plane-wave model of photoemission the measured momentum-dependent photoemission intensity of electrons emitted from  molecular orbital can be expressed as~\cite{Puschnig09sci}
\begin{align}
    I(\mathbf{k}) = \left|\mathbf{A\cdot k}\right|^2 \left| \mathcal{F} \left(\psi(\mathbf{r})\right)\right|^2 \delta\left(E_b + E_{\rm kin} + \Phi - h\nu\right),
    \label{eq:POT}
\end{align}
where $\psi(\mathbf{r})$ is the real-space electronic wavefunction, $\mathcal{F}$ is the Fourier transform, $\left|\mathbf{A\cdot k}\right|^2$ is a polarization factor defined by the vector potential $\mathbf{A}$ of the incoming electromagnetic field. The Dirac $\delta$ function ensures energy conservation of the photoemission process, which includes the photon energy $h\nu$, the electron binding energy $E_b$, the work function $\Phi$ and the kinetic energy $E_{\rm kin}$ of the emitted photoelectron. This model has been successfully applied to analyze orbital wavefunctions of transient excited states in PTCDA~\cite{Wallauer20sci} and extended to the description of the photoemission signature of excitons in C$_{60}$~\cite{Bennecke24natcom}. According to refs.~\cite{Bennecke24natcom, Kern23prb}, the photoemission signature of the hX with multiple hole contributions, but only a single electron contribution, must feature a two-peak structure, where the momentum distribution of both peaks resembles the Fourier transform of the LUMO of PTCDA as described by equation \ref{eq:POT}. Based on this model, we calculate the expected momentum map of the HOMO and the hX considering all the different orientations of the PTCDA molecule~\cite{ZhangNanoResearch2022}. The real-space molecular orbitals calculated by DFT are extracted from reference~\cite{puschnig2020organic}. The results are plotted in extended Fig.~\ref{extFig_POT}c,g. 

\vspace{1cm}
\noindent\textbf{Calculation of the electronic structure}
\vspace{0.2cm}

A \GW{} treatment of the herringbone-type WSe$_2$/PTCDA heterostructure is beyond current computational possibilities. Instead, to meet the experimental conditions as closely as possible, we consider a configuration of a PTCDA molecule adsorbed on a 4$\times$4$\times$1 supercell of the pristine WSe$_2$ structure with an in-plane lattice parameter of 3.317 \r{A}~ (extended Fig.~\ref{extFig_calcBandstructure}). We optimize the atomic structure, consisting of 86 atoms, using the all-electron code FHI-aims~\cite{Blum2009} by minimizing the amplitude of the interatomic forces below a threshold value of $10^{-3}$ eV \r{A}$^{-1}$. For all species a \textit{tight} basis is used. The resulting adsorption geometry is shown in extended Fig.~\ref{extFig_calcBandstructure}a. The PTCDA molecule is slightly tilted, with the shortest and longest distance to the substrate being 2.87 \r{A} and 4.98 \r{A}, respectively, measured from the top of the substrate.

The ground-state, \GW, and BSE calculations are performed using the all electron full-potential code \exciting{}~\cite{Gulans2014}, which implements the family of linearized augmented plane wave plus local orbitals (LAPW+LO) methods. The muffin-tin (MT) spheres of the inorganic component are chosen to have equal radii of 2.2 bohr. For PTCDA, the radii are 0.9 bohr for hydrogen (H), 1.1 bohr for carbon (C), and 1.2 for oxygen (O). The electronic properties are calculated first using DFT with the generalized gradient approximation in the Perdew-Burke-Ernzerhof (PBE) parametrization for the exchange-correlation functional. The sampling of the Brillouin zone (BZ) is carried out with a homogeneous 3$\times$3$\times$1 Monkhorst-Pack $\mathbf{k}$-point grid. To account for van der Waals forces and intermolecular interactions, we adopt the Tkatchenko-Scheffler (TS) method~\cite{Tkatchenko2009}. The quasi-particle (QP) energies are computed within the \GW~approximation~\cite{Nabok2016}, where we include 200 empty states to compute the frequency-dependent dielectric screening within the random-phase approximation. A 2D truncation of the Coulomb potential in the out-of-plane direction $z$ is employed~\cite{Fu2016}. The band structure is computed by using interpolation with maximally-localized Wannier functions~\cite{Tillack2020} and Fourier interpolation (cf. extended Fig.~\ref{extFig_calcBandstructure}b and \ref{extFig_calcBandstructure}c, respectively).

To allow a direct comparison with the experimental ARPES data (Fig.~1d), we unfold the theoretical band structure by symmetry mapping of the Bloch-vector-dependent quantities defined in the supercell into the unit-cell calculations. Here, the wavefunctions are constructed in a uniform real-space grid of 120$\times$120$\times$120 and used to calculate the spectral function (Fig.~1f). 

The quasi-particle band gap of WSe$_2$ in the heterostructure is in good agreement with that measured by STS~\cite{Guo22nanoresearch}, however, the PTCDA gap is underestimated, which is most evident in the level alignment of the HOMO. This discrepancy can be explained by the PBE starting point, which tends to underestimate the band gap of the organic side more than that of the TMD. As we can deduce from calculations for pyrene on MoS$_2$~\cite{GonzalezOliva2024}, GW@HSE would increase the gaps on both sides of the interface, however, to a different extent. Like in pyrene on MoS$_2$, we do not expect a qualitative change in the level alignment. To overcome this mismatch, we apply a scissors shift to the molecular levels. Shifting the LUMO by -50 meV closer to the experimental value~\cite{Guo22nanoresearch} results in very good agreement of the excitonic spectrum with experiment (see below).

\vspace{1cm}
\noindent\textbf{Calculation of the exciton spectrum} 
\vspace{0.2cm}

For the calculation of the exciton spectrum, we solve the Bethe-Salpeter equation (BSE) on top of the QP band structure, where the screened Coulomb potential is computed using 100 empty bands. In the construction and diagonalization of the BSE Hamiltonian, 16 occupied and 14 unoccupied bands are included, and a 12$\times$12$\times$1 shifted $\mathbf{k}$-point mesh is adopted. Calculations are performed using the BSE module~\cite{Vorwerk2019} of the \exciting{} code.

\vspace{1cm}
\noindent\textbf{Calculation of the correlation function}
\vspace{0.2cm}

Following the definition in ref.~\cite{Sharifzadeh13jpcl}, we calculate the electron-hole correlation function
\begin{align}
    F^{i} (r) = \int_{\Omega} d^3 \mathbf{r}_e | \psi_{i} (\mathbf{r}_h = \mathbf{r}_e + \mathbf{r} , \mathbf{r}_e) |^2 
\end{align}

\noindent where $F^{i}$ describes the probability of finding electron and hole separated by the vector $\mathbf{r} = \mathbf{r}_h - \mathbf{r}_e$. 
We approximate this integral by a discrete sum over a finite number of fixed electron coordinates. For each electron position, the hole probability $|\psi_{i} (\mathbf{r}_h = \mathbf{r}_e + \mathbf{r} , \mathbf{r}_e) |^2$ is computed on an evenly spaced, dense grid of 100$\times$100$\times$100 sampling points, covering approximately 3$\times$3$\times$1 supercells.
For the hX, we sampled 60 positions on the PTCDA molecule (0.5 \r{A}$^{-1}$ below and above the carbon and oxygen atoms) since its electronic contribution is almost entirely comprised of the LUMO of PTCDA. Similarly, we calculated the electron-hole correlation function of the K-exciton (cf. extended Fig.~\ref{extFig_Aexciton_spatial}), which is completely localized in the WSe$_2$ layer. Here, we sampled 16 positions close to the W atoms where we expect a high probability of finding the electron.

For further analysis, the three dimensional correlation function is splitted into its in-plane and out-of-plane components (Fig.~4, extended Fig.~\ref{extFig_Aexciton_spatial}). Notably, the in-plane component shows a distinct periodic pattern (insets Fig.~4b, extended Fig.~\ref{extFig_Aexciton_spatial}b). Thus, to extract the in-plane radial profile and the root mean square (RMS) radius, we first filter the data in Fourier space thereby smoothing it in the real-space. For the hX, we extract the Wannier-like and Frenkel-like in-plane distribution by only integrating the correlation function over the inter- or intralayer contribution, respectively.

\vspace{1cm}
\noindent\textbf{Spatial analysis of the K-exciton}
\vspace{0.2cm}

In analogy the analysis of the spatial structure of the hX exciton in Fig.~4, we analyze the K-exciton wavefunction (extended Fig.~7). From the BSE calculation, we find that $F^{\rm K}(r_\perp)$ is dominated by a single peaked feature centered around $r_\perp\approx0$~\r{A}, implying that the exciton is of pure intralayer character (extended Fig.~\ref{extFig_Aexciton_spatial}a). Consequentially, the probability density to find K-excitons in the WSe$_2$ layer is nearly 100\%. Consistent with this, the two smaller side peaks located at a distance corresponding to the distance between the tungsten and selenium planes d$_{\rm W,Se}$, can be attributed to a residual probability of the electron or/and hole being at the selenium atoms.

Due to its hydrogen-like structure, the in-plane electron-hole probability distribution of the K-exciton can directly be reconstructed from the experimental photoemission momentum fingerprint via Fourier analysis~\cite{Man21sciadv, dong20naturalsciences, Schmitt22nat, Karni22nat}. 
This allows a direct comparison of $F^{\rm K}(\ve{r}_{\parallel})$ between theory and experiment (extended Fig.~\ref{extFig_Aexciton_spatial}b). Both, theory and experiment, confirm the pure Wannier-like character of the K-exciton since the radial distribution is much larger than the WSe$_2$ lattice constant. For a more quantitative analysis, we compare the extracted root mean square (RMS) radii to be (r$_{\rm K}^{\rm BSE}=12$~\r{A}) (theory) and r$_{\rm K}^{\rm exp}=10\pm 1$~\r{A} (experiment), which are in excellent agreement. Note that due to a finite momentum resolution of the photoemission signal the derived RMS radius is a lower limit of the true value. 

For comparison, the calculated RMS radii of the K-exciton and both components of the hX-exciton are summarized in table \ref{table:RMS}.

\begin{table}[h]
\caption{Theoretical and experimental extracted RMS radii of the in-plane relative probability distribution of the K-exciton and the hX.}
\begin{tabular}{cccccc}
 \multicolumn{2}{c}{K (\AA)} & & \multicolumn{3}{c}{hX (\AA)} \\ 
 \cline{0-1} \cline{4-6}
 Experiment & Theory & & Total & Frenkel & Wannier \\  \cline{0-1} \cline{4-6}
 $10 \pm 1$ &  12 & & 7 & 6 & 10  \\
\end{tabular}
\label{table:RMS}
\end{table}

The distinct intralayer and Wannier-like character of the K-exciton can be further visualized by plotting the isosurface of representative fixed electron and hole position (extended Fig.~\ref{extFig_Aexciton_spatial}c) which stands in clear contrast to the isosurfaces of the hX (Fig.~4).


\section{ACKNOWLEDGEMENTS}

The authors thank Christian Kern for valuable discussions.

The Göttingen team is funded by the Deutsche Forschungsgemeinschaft (DFG, German Research Foundation) - 432680300/SFB 1456 (project B01), 217133147/SFB 1073 (projects B07 and B10), and 535247173/SPP2244. C.D. and I.G.O. acknowledge support from the Deutsche Forschungsgemeinschaft (DFG, German Research Foundation) via SFB 951 (project 182087777) and the European Union's Horizon 2020 research and innovation program under the agreement No. 951786 (NOMAD CoE). I.G.O thanks the DAAD (Deutscher Akademischer Austauschdienst) for financial support and acknowledges fruitful discussions with Sebastian Tillack and Benedikt Mauer. Computing time on the supercomputers Lise and Emmy at NHR@ZIB and NHR@Göttingen is gratefully acknowledged. P.P. acknowledges support from the Austrian Science Fund (FWF) project I 4145 and from the European Research Council (ERC) Synergy Grant, Project ID 101071259.  K.W. and T.T. acknowledge support from the JSPS KAKENHI (Grant Numbers 21H05233 and 23H02052) and World Premier International Research Center Initiative (WPI), MEXT, Japan.

\section{DISCLOSURE STATEMENT}

No potential conflict of interest was reported by the author(s).

\section{Data availability}
The experimental data and evaluation scripts will be made available on GRO.data. The theoretical data can be downloaded under the following link  \url{https://dx.doi.org/10.17172/NOMAD/2024.10.11-1}.

\section{Authors contributions}
D.St., R.T.W., P.P., C.D., G.S.M.J., M.R. and S.M. conceived the research. W.B., P.W. and A.S. fabricated the sample. W.B., J.P.B., P.W., D.S. and M.M. carried out the time-resolved momentum microscopy experiments. W.B. analyzed the experimental data. I.G.O. carried out the theoretical calculations guided by C.D. All authors discussed the results. G.S.M.J., M.R. and S.M. were  responsible for the overall project direction. W.B., I.G.O., G.S.M.J., M.R. and S.M. wrote the manuscript with contributions from all co-authors. K.W. and T.T. synthesized the hBN crystals.

\clearpage

\bibliography{bibtexfile}

\clearpage \newpage

\renewcommand{\figurename}{Extended Fig.}
\def\bibsection{\section*{\refname}} 
\setcounter{figure}{0} 

\begin{figure*}[hbt!]
    \centering
    \includegraphics[width=.5625\linewidth]{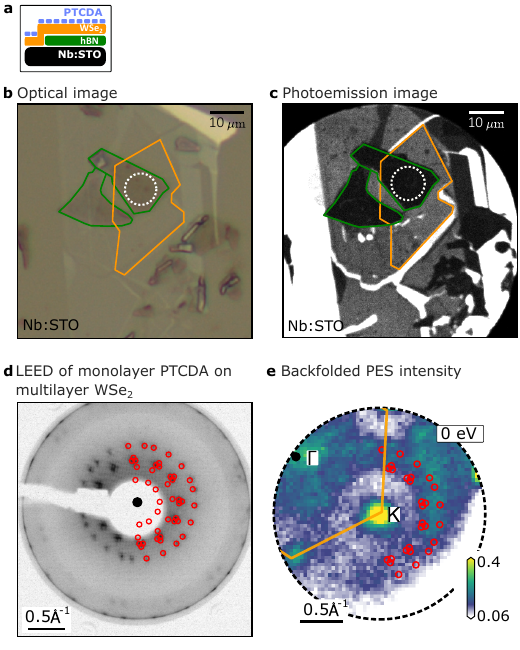}
    \caption{\textbf{Analysis of the real-space structure of the WSe$_2$/PTCDA heterostructure.} 
    \textbf{a} Sketch of the layered sample structure. \textbf{b}, \textbf{c} Optical microscope and photoemission real-space image of the sample before PTCDA evaporation. The different flakes are marked with the same colors as used in \textbf{a}. Note that the marked WSe$_2$ area corresponds to an intact monolayer (without cracks) whereas the complete monolayer as seen by the contrast in \textbf{b} and \textbf{c} was larger. \textbf{d} LEED pattern of monolayer PTCDA on multilayer WSe$_2$ recorded with a beam energy of 24~eV. The red circles correspond the superstructure defined by the matrix 
    $M = \begin{bmatrix}
    1.58       & 6.78 \\
    4.39       & 1.32 
    \end{bmatrix}$. \textbf{e} The presence of a well-ordered PTCDA monolayer on the WSe$_2$ monolayer is confirmed by the appearance of umklapp-scattering replicas of the WSe$_2$ band structure, here shown for the valence band maximum at the K point (red circles). The replicas can be directly compared to the LEED pattern in \textbf{d}.}
    \label{extFig_SamplePrep}
\end{figure*}

\begin{figure*}[hbt!]
    \centering
    \includegraphics[width=\linewidth]{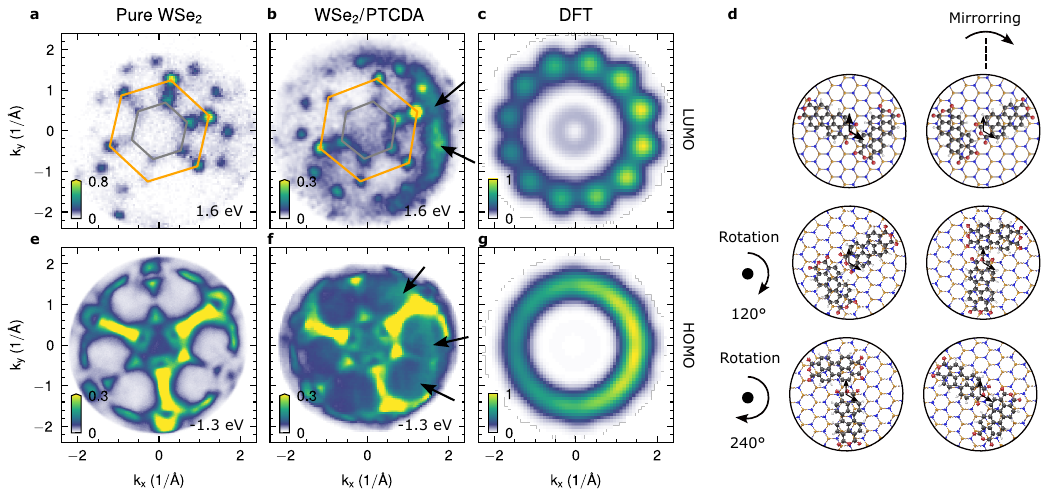}
    \caption{\textbf{Direct comparison of ARPES data collected on pristine monolayer WSe$_2$ and the WSe$_2$/PTCDA heterostructure.}
    \textbf{a,b,e,f} Momentum-maps collected at energies of the valence bands (bottom row) and at energies of excitonic photoemission signatures (top row). Photoemission signatures of the HOMO and the hX are labeled by arrows in \textbf{f} and \textbf{b}, respectively.
    \textbf{c,g} Simulated momentum maps from DFT calculations of the LUMO and HOMO of PTCDA using the plane wave model of photoemission~\cite{Puschnig09sci} and accounting for the herringbone structure and the different mirror and rotational domains shown in \textbf{d}. The DFT data are extracted from reference~\cite{puschnig2020organic}.}
    \label{extFig_POT}
\end{figure*}

\begin{figure*}[hbt!]
    \centering
    \includegraphics[width=\linewidth]{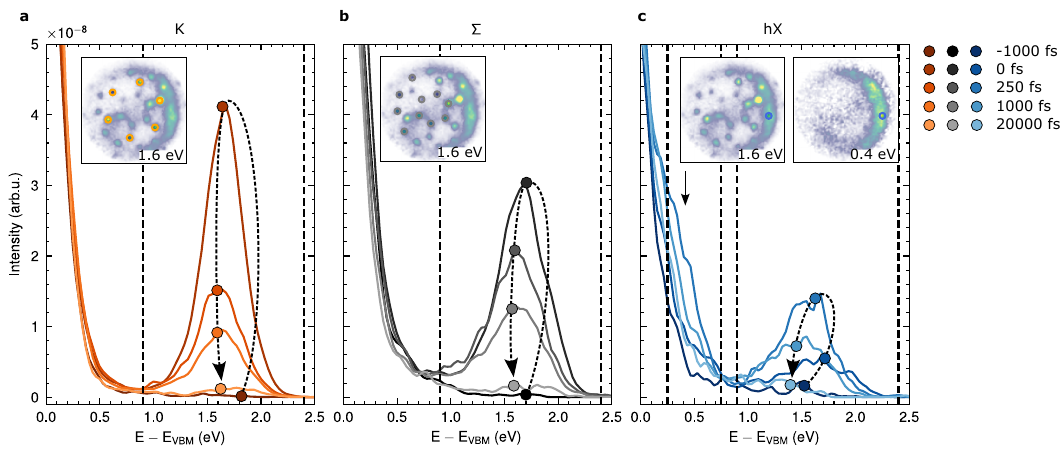}
    \caption{\textbf{Momentum-filtered energy-distribution-curves of the K-exciton (\textbf{a}), the $\Sigma$-exciton (\textbf{b}), and the hX (\textbf{c}).} The dashed vertical lines indicate the energy range used to extract the delay-dependent data shown in Fig.~5. \textbf{c} The vertical black arrow highlights the appearance of the lower energy photoemission maximum that is attributed to the hX photoemission signature.}
    \label{extFig_EDCs}
\end{figure*}

\begin{figure*}[hbt!]
    \centering
    \includegraphics[width=\linewidth]{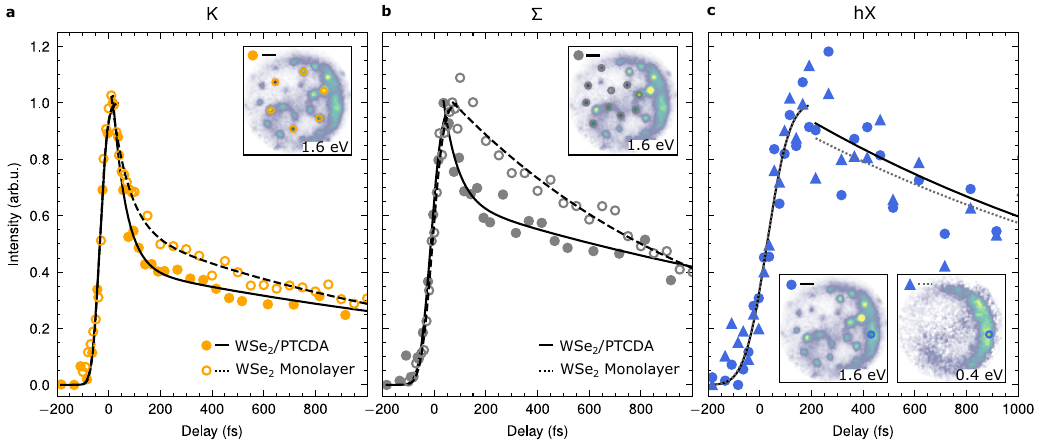}
    \caption{\textbf{Exciton formation and relaxation dynamics of the bright K-exciton (\textbf{a}), the momentum dark $\Sigma$-exciton (\textbf{b}) and  the hX (\textbf{c}).} The respective regions of interest in momentum space are shown in the insets. The time-dependent data collected on WSe$_2$/PTCDA (filled symbols) is directly plotted next to data collected on pristine monolayer WSe$_2$ (open symbols; data taken from ref.~\cite{Bange232DMaterials}). The hX double-peak photoemission structure is evaluated separately in energy ranges of 0.9-2.4~eV (circle) and 0.3-0.7~eV (triangles). The rise and decay dynamics are fitted with an error and a (bi-) exponential function, respectively (cf. methods). The fit parameters are summarized in Table~1.}
    \label{extFig_tempAnalysis}
\end{figure*}

\begin{figure*}[hbt!]
    \centering
    \includegraphics[width=\linewidth]{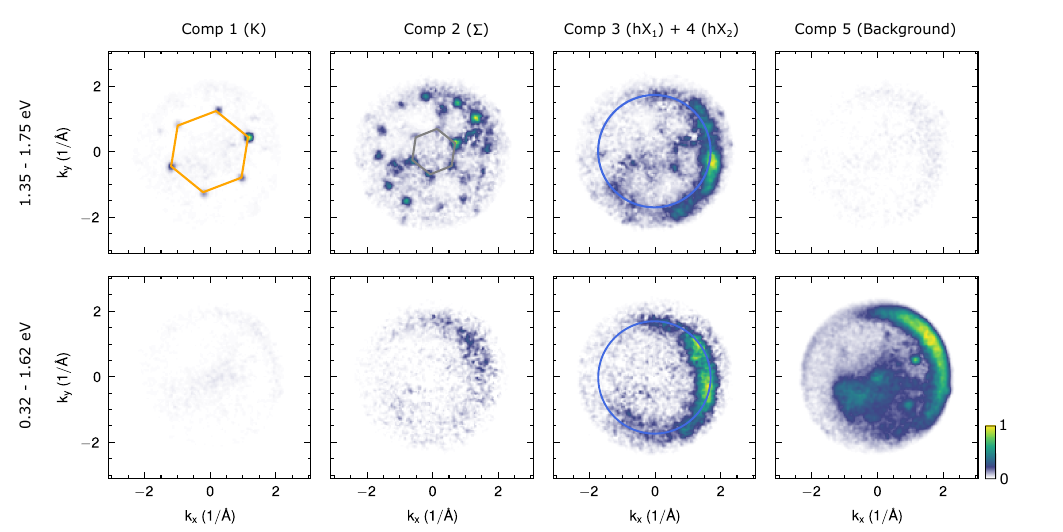}
    \caption{\textbf{NMF component analysis of the ARPES data.}
    Components 1, 2, 3, and 4 follow the temporal dynamics extracted for the K, $\Sigma$, hX$_1$ and hX$_2$ photoemission features, respectively. Comp 5 corresponds to the static component, \ie the time-independent background which is subtracted from the raw data for the momentum-maps and EDCs shown in Fig.~2. 
    The photoemission features of the K-excitons, $\Sigma$-excitons and the hX are marked by an orange hexagon, a gray hexagon and a blue circle, respectively, in the components where most prominent. The energy region of interests are chosen to be the same as the one shown in Fig.~2. Each component is normalized to the maximal value of the upper and lower panel.}
    \label{extFig_NMF}
\end{figure*}

\begin{figure*}[hbt!]
    \centering
    \includegraphics[width=\linewidth]{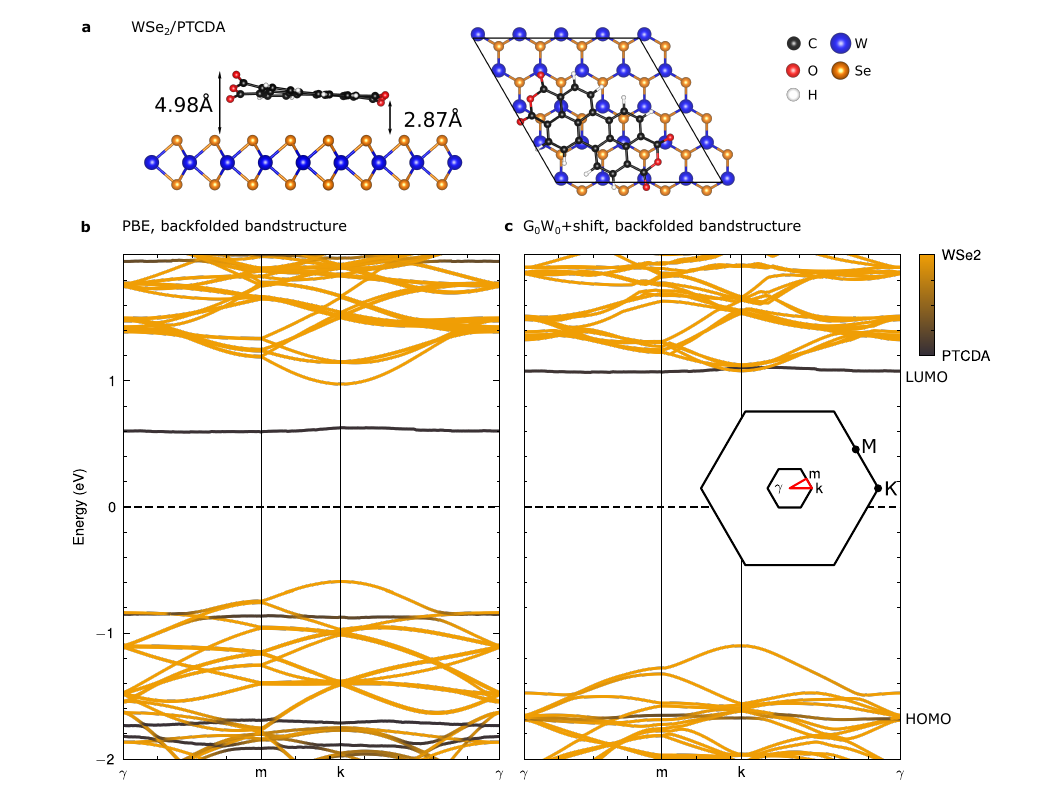}
    \caption{\textbf{Calculated supercell and single-particle band structure.}
    \textbf{a} Optimized structure of PTCDA adsorbed on WSe$_2$ monolayer in a 4$\times$4$\times$1 WSe$_2$ supercell. 
    \textbf{b,c} Band structure of WSe$_2$/PTCDA obtained by PBE and G$_0$W$_0$ calculation analyzed along the directions of the backfolded Brillouin zone indicated in the inset of \textbf{c}. The bands are colored according to their PTCDA (black) or WSe$_2$ (orange) character. The G$_0$W$_0$ results in \textbf{c} already include the scissors shift applied to the LUMO according to existing STS data in literature~\cite{Zheng16acsnano, Guo22nanoresearch}.}
    \label{extFig_calcBandstructure}
\end{figure*}

\begin{figure*}[hbt!]
    \centering
    \includegraphics[width=\linewidth]{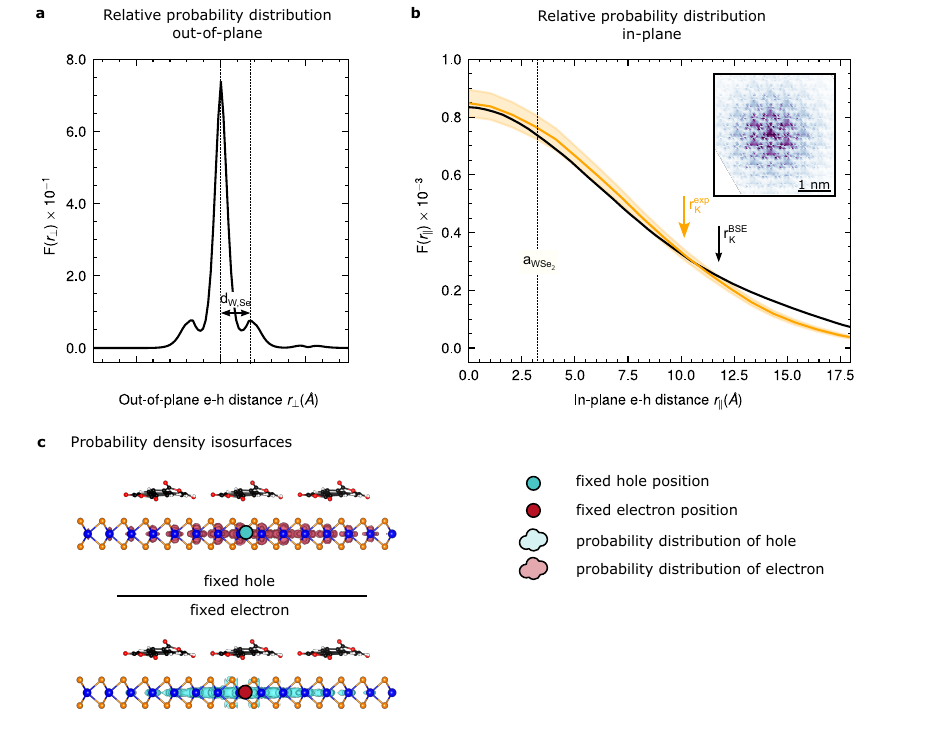}
    \caption{\textbf{Analysis of the real-space properties of the K-exciton}. The relative out-of-plane \textbf{a} and in-plane components \textbf{b} of the electron-hole correlation function $F^i(\ve{r}) = F(\ve{r}_e-\ve{r}_h)$ are analyzed. 
    \textbf{a} The out-of-plane (r$_\perp$) component directly confirms the intralayer nature ($r_\perp\approx 0$~\r{A}). Next to the dominant peak at $r_\perp=0$~\r{A}, two side peaks are present at a distance that matches the distance between the tungsten and selenium planes (d$_{\rm W,Se}$) of the WSe$_2$ monolayer. 
    \textbf{b} The in-plane component can be accessed by theory (black) and experiment (orange). Here, the experimental distribution corresponds to the weighted mean from the angular-averaged probability distribution of all 6 K-points. The extracted RMS radius is marked by arrows. The inset corresponds to the two-dimensional representation of  $F^{\rm K}(\ve{r}_{\parallel})$. \textbf{c} Exemplary probability density isosurfaces of the K-exciton for fixed hole (top) and fixed electron (bottom) position confirming the pure intralayer and Wannier-like character of the K-exciton.} 
    \label{extFig_Aexciton_spatial}
\end{figure*}

\end{document}